\documentclass[useAMS,usenatbib]{mn2e}

\title[On the origin of the circular polarization in radio pulsars]
{On the origin of the circular polarization in radio pulsars}
\author[G. Gogoberidze \& G. Z. Machabeli]{G. Gogoberidze$^{1}$\thanks{E-mail:
gogober@geo.net.ge} and  G. Z. Machabeli$^{1}$\\
$^{1}$ Abastumani Astrophysical Observatory, Al. Kazbegi Avenue 2a,
0160 Tbilisi, Georgia}

\begin{document}

\date{}

\pagerange{\pageref{firstpage}--\pageref{lastpage}} \pubyear{2004}

\maketitle

\label{firstpage}

\begin{abstract}
Properties of circularly polarized waves are studied in the pulsar
magnetosphere plasma. It is shown that some observational
characteristics of the circular polarization observed in the pulsar
radio emission can be qualitatively explained in the framework of
the model based on anomalous Doppler resonance. Performed analysis
provides that if the difference between Lorentz factors of electrons
and positrons is relatively high, one of the circularly polarized
waves becomes super-luminal and therefore can not be generated by
cyclotron instability. We suggest that this case corresponds to the
pulsars with the domination of one hand of circular polarization
through the whole averaged pulse profile at all observed
frequencies. For intermediate values of the difference between
Lorentz factors both circularly polarized waves are generated, but
the waves of one handness are much more effectively generated for
high frequencies, whereas generation of another handness dominates
for low frequencies. This should correspond to the pulsars with
strong frequency dependence of the degree of circular polarization.
The case of relatively small difference between Lorentz factors
corresponds to the pulsars with sign reversal of the circular
polarization in the centre of averaged pulse profiles.
\end{abstract}


\section{Introduction}

Radio emission from pulsars is highly polarized. In some cases the
fraction of linear polarization in the total radio emission is
close to 100 \%. On the other hand, the circular polarization $V$
is the highest among observed natural sources of electromagnetic
radiation. Observations of circular polarization of individual
pulses \citep{CS69,MTH75,HMXQ98,KJMLE03,KJ04} show high degree of
$V$, usually several tens of per cent and very irregular structure
of circular polarization through the pulse.

Significant circular polarization is also usually observed in the
average pulse profiles of most pulsars. However, as a rule
averaged profiles have much smaller degree of circular
polarization then individual pulses. It tends to peak near the
centre of the averaged pulse profile. In many pulsars the circular
polarization reverses its sign near the centre, while in other
pulsars the same sign of $V$ retained throughout. According to
\citet{RR90} these two types of circular polarization behavior are
called 'antisymmetric' and 'symmetric' types respectively. The
pulsars with the symmetric circular polarization often show strong
frequency dependence of the degree of circular polarization.

In spite of extensive observational data even the origin of high
circular polarization in pulsar radio emission remains mainly
unclear. Usually it is supposed that the circular polarization is
related to either the pulsar radio emission mechanism or propagation
effects in the pulsar magnetosphere \citep{CR79,ML04}. \citet{CR79}
suggested that asymmetry between electrons and positrons in the
magnetosphere plasma converts linear polarization into circular one
during electromagnetic wave propagation in the magnetosphere.
According to another model \citep{KMM91}, cyclotron instability is
responsible for the observed circular polarization in pulsar radio
emission. As it was shown by \citet{KMM91}, due to the relative
motion of electrons and positrons that was supposed by \citet{CR77},
two circularly polarized electromagnetic waves can exist in the
pulsar magnetosphere for relatively small angles of propagation with
respect to the pulsar magnetic field. According to this model these
circularly polarized waves are generated by cyclotron instability.
In the framework of this model, the antisymmetric type circular
polarization is a consequence of relative drift motion of electrons
and positrons in the curved magnetic field. In general it seems that
both generation and propagation effects should contribute to the
observed circular polarization of pulsars.

In the presented paper we study properties of circularly polarized
waves in the magnetosphere plasma and ignore propagation effects.
Performed analysis provides that if the difference between Lorentz
factors of electrons and positrons is relatively high, one of the
circularly polarized waves becomes super-luminal and therefore can
not be generated by kinetic mechanisms, such as cyclotron
instability. We suggest that this case corresponds to the pulsars
with symmetric circular polarization profiles that shows the
domination of one handness circular polarization at all observed
frequencies. For intermediate values of the difference between
Lorentz factors both of circularly polarized waves are generated,
but the waves of one handness are much more effectively generated
for high frequencies, whereas generation of another handness
dominates for low frequencies. This should correspond to the pulsars
with symmetric circular polarization profiles that shows strong
frequency dependence of the degree of circular polarization. The
case of relatively small difference between Lorentz factors was
discussed by \citet{KMM91} and corresponds to the pulsars with
antisymmetric circular polarization profiles.

The paper is organized as follows. Properties of circularly
polarized waves in the magnetosphere plasma are discussed in
section 2. Different regimes of cyclotron instability development
are studied in  section 3. The conclusions are summarized in
section 4.

\section{Circularly polarized waves in the magnetosphere plasma}

It is generally assumed that the pulsar magnetosphere is filled by
dense relativistic electron-positron plasma flowing along the open
magnetic field lines, which is generated as a consequence of the
avalanche process first described by \citet{GJ69} and developed by
\citet{S71}. This plasma is multi-component (see, e.g., Arons 1981),
with a one-dimensional distribution function, containing: (i)
electrons and positrons of the bulk of plasma with mean Lorentz
factor of $\gamma _{p}$ and density $n_p$; (ii) particles of the
high-energy `tail' of the distribution function with $\gamma _{t}$
and $n_{t}$, stretched in the direction of positive momenta; (iii)
the ultrarelativistic ($\gamma_{b}\sim 10^{6}$) primary beam with so
called `Goldreich-Julian' density $n_b \approx 7\times
10^{-2}B_{0}P^{-1}(R_{0}/r)^{3}\,{\rm cm}^{-3}$ (where $P$ is a
pulsar period, $R_{0}$ is a neutron star radius, $B_{0}$ is a
magnetic field value at the stellar surface and $r$ is a distance
from the neutron star's centre), which is much less than $n_{p}$
($\kappa \equiv n_{p}/n_{b}\gg1$). The one dimensional distribution
function of electrons $f_-$ and positrons $f_+$ are shifted with
respect to each other due to the presence of the primary beam and
quasi neutrality of the plasma. The value
\begin{equation}
\Delta \gamma = \gamma_+ -\gamma_-=\int f_+\gamma d p_\parallel
-\int f_-\gamma dp_\parallel, \label{eq:21}
\end{equation}
is small but of decisive significance in explaining of the
polarization properties of pulsars. The distribution functions are
normalized such that $\int f_\pm d p_\parallel \equiv 1$.

For clarity we assume that the primary beam consists of particles
with positive charge. In this case according to \citet{CR77}
\begin{equation}
\Delta \gamma \approx \gamma_p^4/\gamma_b >0. \label{eq:22}
\end{equation}

An extensive analysis have been conducted \citep{VKM85,AB86,KMM91}
in order to study the dispersion properties of the waves propagating
through the highly magnetized relativistic electron-positron plasma
of pulsar magnetosphere. If the $z$ axis is directed along the
pulsar magnetic field and the wave vector ${\bf k}$ is assumed to
lie in the $(y,z)$ plane, and the drift motion of the plasma
particles due to the curvature of the magnetic field is neglected,
the components of permittivity tensor are \citep{KMM91}:

\begin{equation}
\epsilon_{xx}=\epsilon_{yy}=1-\frac{1}{2}\sum_\alpha\frac{\omega_{p\alpha}^2}{\omega^2}
\int \frac{dp_{z}}{\gamma}(\omega-k_zv_z)A^+_\alpha f_\alpha,
\label{eq:23}
\end{equation}
\begin{equation}
\epsilon_{zz}=1-\sum_\alpha\omega_{p\alpha}^2 \int
\frac{dp_{z}f_\alpha}{\gamma^3(\omega-k_zv_z)^2}, \label{eq:24}
\end{equation}
\begin{equation}
\epsilon_{yx}=-\epsilon_{xy}=\frac{i}{2}\sum_\alpha\frac{\omega_{p\alpha}^2}{\omega^2}
\int \frac{dp_{z}}{\gamma}(\omega-k_zv_z)A^-_\alpha f_\alpha,
\label{eq:25}
\end{equation}
\begin{equation}
\epsilon_{zx}=-\epsilon_{xz}=\frac{i}{2}\sum_\alpha\frac{\omega_{p\alpha}^2}{\omega^2}
\int \frac{dp_{z}}{\gamma}k_yv_z A^-_\alpha f_\alpha,
\label{eq:26}
\end{equation}
\begin{equation}
\epsilon_{yz}=\epsilon_{zy}=-\frac{i}{2}\sum_\alpha\frac{\omega_{p\alpha}^2}{\omega^2}
\int \frac{dp_{z}}{\gamma}k_yv_z A^+_\alpha f_\alpha,
\label{eq:27}
\end{equation}
where
\begin{equation}
A_\alpha^\pm \equiv \frac{1}{\Omega_\alpha^-} \mp
\frac{1}{\Omega_\alpha^+},~~\Omega_\alpha^\pm \equiv
\omega-k_zv_z\pm \frac{\omega_{B\alpha}}{\gamma}, \label{eq:28}
\end{equation}
and
\begin{equation}
\omega_{p\alpha}^2\equiv \frac{4\pi e^2 n_\alpha}{m},
~~\omega_{B\alpha} \equiv \frac{e_\alpha B}{mc} \label{eq:29}
\end{equation}
are plasma and cyclotron frequencies of corresponding kind of
particles respectively.

Introducing the system with $z^\prime$ axis directed along ${\bf
k}$ the equations for waves in the plasma takes the form:
\begin{equation}
\left[ (k^{\prime 2}\delta_{lm}-k^\prime_l
k^\prime_m)\frac{c^2}{\omega^2} -\epsilon_{lm}^\prime \right]
E^\prime_m=0. \label{eq:210}
\end{equation}

The radio emission is thought to be due to one of several possible
instabilities in which the distribution of particles causes waves in
a specific wave mode to grow. These waves then propagate in the pair
plasma of a pulsar magnetosphere, transform into vacuum-like
electromagnetic waves as the plasma density drops, enter the
interstellar medium, and reach an observer as the pulsar radio
emission.

In the general case of oblique propagation with respect to the
magnetic field three different wave modes can be distinguished.
One of these modes is the super-luminous O-mode. Identification of
the other modes depends on the angle $\theta$ between ${\bmath k}$
and ${\bmath B}$, where ${\bmath k}$ a wave-vector and ${\bmath
B}$ is a local magnetic field. If
\begin{equation}
\theta^2 \gg \frac{\omega}{\omega_B} \frac{\Delta
\gamma}{\gamma_p^4} \equiv \theta_0^2, \label{eq:211}
\end{equation}
where $\omega_B\equiv |e|B/mc$, there exists two linearly
polarized waves: the purely electromagnetic X-mode and the
sub-luminous Alfv\'{e}n (A) mode. The A~mode, as well as O-mode,
are of mixed electrostatic-electromagnetic nature. Electric field
vectors of the O and A-modes lie in the $\left( {\bmath
k}\,{\bmath B} \right)$ plane, while the electric field of the
X-mode perpendicular to this plane.

In the opposite limiting case
\begin{equation}
\theta^2 \ll \theta_0^2, \label{eq:212}
\end{equation}
there exist two circularly polarized waves with left-handed and
right-handed polarization. In the case of propagation along ${\bmath
B}$, according to equations (\ref{eq:23})-(\ref{eq:27}) and
(\ref{eq:210}) dispersion equations for these waves are:
\begin{equation}
\frac{k^2c^2}{\omega^2}=\epsilon_{xx}\pm i\epsilon_{xy}=
1-\sum_\alpha\frac{\omega_{p\alpha}^2}{\omega^2} \int
\frac{dp_{z}}{\gamma}\frac{\omega-k v}{\Omega^\pm_\alpha}
f_\alpha. \label{eq:213}
\end{equation}
Assuming $\omega/2\gamma |\omega_{B\alpha}| \ll 1$ and expanding
the integrand in the series with respect to this small parameter
we obtain:
\begin{equation}
\omega_{1,2}=kc(1-\delta) \pm \Delta\gamma\omega_B\delta,
\label{eq:214}
\end{equation}
where $\delta\equiv \omega_{p}^2/4\gamma_p^3\omega_B^2$ and upper
sign corresponds to right handed circularly polarized wave.

Analysis of cyclotron resonance condition
\begin{equation}
\omega-k_zv_z \pm \frac{\omega_{B\alpha}}{\gamma}=0
\label{eq:213a}
\end{equation}
provides \citep{MU79}, that the wave are generated only at
anomalous Doppler effect [i.e., when the third term in equation
(\ref{eq:213a}) is positive]. Consequently, left and right-handed
circularly polarized waves are excited by resonant positrons and
electrons respectively. It can be readily shown, that only the
particles of the tale and primary beam are involved in this
process. The growth rate of the instability is
\begin{equation}
\Gamma \approx \frac{\pi}{2}
\frac{\omega_{p,res}^2}{\omega_0\gamma_{T,res}}, \label{eq:215}
\end{equation}
where $\omega_0$ is the resonant frequency
\begin{equation}
\omega_0 \approx \frac{\omega_B}{\delta \gamma_{res}},
\label{eq:216}
\end{equation}
and $\gamma_{T,res}$ and $\gamma_{res}$ are thermal spread and
average Lorentz factors of the resonant particles respectively.
Cyclotron resonance with the particles of the bulk of plasma
causes the damping of the waves with the decrement
\begin{equation}
\Gamma^\prime \approx -\frac{\pi}{2}
\frac{\omega_{p}^2}{\omega_0^\prime \gamma_{p}}, \label{eq:217}
\end{equation}
where the frequency of the damped waves $\omega_0^\prime \sim
2\gamma_p \omega_B$.

\citet{KMM91} suggested that the 'core' component of pulsar
emission is generated via the cyclotron instability.
Electromagnetic waves are generated at the distance of several
hundred neutron star radii above the surface. In the framework of
this model the sign reversal of the circular polarization,
frequently observed near the centre of mean pulse profile, is
caused by drift motion of electrons and positrons in the curved
magnetic field. As it was mentioned above, the effects of the
curvature drift motion of the particles are not considered in the
presented paper.

In the next section we will show, that main properties of the two
groups of pulsars, that have high circular polarization component
either of one sign for the whole frequency range or different sign
of circular polarization for different frequencies, can also be
naturally explained in the framework of this model.

\section{Different regimes of the cyclotron instability development}

As it was mentioned above, the electromagnetic waves can be
generated only by cyclotron resonance at anomalous Doppler effect.
As it can be readily seen, this means that only the generation of
sub-luminal waves is possible. Equation (\ref{eq:214}) shows, that
in the presence of the relative motion of electrons and positrons
($\Delta\gamma\neq 0$) the waves in the magnetosphere can be
sub-luminal as well as super-luminal. If $kc<\Delta \gamma
\omega_B$, the right-handed circularly polarized wave becomes
super-luminal and therefore can not be generated by cyclotron
instability (note, that if $\Delta \gamma$ defined by equation
(\ref{eq:21}) is negative, then the same is true for left-handed
circularly polarized wave). Consequently, the development of the
cyclotron instability strongly depends on parameter
\begin{equation}
\Upsilon \equiv \frac{\Delta \gamma}{2\gamma_p}
\frac{2\gamma_p\omega_B}{\omega}. \label{eq:32}
\end{equation}

The second multiplier in the right hand side of this equation
$2\gamma_p\omega_B/\omega \gg 1$, according to assumption made for
derivation of dispersion relation (\ref{eq:214}). Estimation of the
first multiplier $\Delta\gamma/\gamma_p$ is much more complicated.
According to \citet{RS75} model
\begin{equation}
\gamma_b \approx 3 \cdot 10^6 \left( \frac{B_0}{10^{12}~\rm G}
\right)^{-1/7} \left( \frac{P}{1~\rm sec} \right)^{-1/7} \left(
\frac{R_c}{R_0} \right)^{4/7}, \label{eq:32_1}
\end{equation}
where $R_c$ is curvature radius of the magnetic field lines at the
surface of the pulsar. If the magnetic field of pulsar is strongly
dipole and $R_c \sim R_0$, then $\gamma_p \sim 10^2$ and equation
(\ref{eq:22}) yields $\Delta\gamma/\gamma_p\sim 1$. On the other
hand, \citet{MU79} shown that if the magnetic field of the pulsar
has significant quadrupole component at the surface of the pulsar,
then
\begin{equation}
\gamma_p \approx 3  \frac{R_c}{R_0} \left[ 1+ \left( \frac{B_0}{2
\cdot 10^{12}~\rm G} \right)^2 \right]^{-1/2}, \label{eq:32_2}
\end{equation}
and for $R_c \sim R_0$ we obtain $\Delta\gamma/\gamma_p\sim
10^{-4}$. Consequently, in principle $\Upsilon$ can vary in the wide
range from $10^{-3}$ to $10^2$ and different regimes of the
cyclotron instability development should be considered separately.

\subsection{The case $\Upsilon \ll 1$}

Substituting equation (\ref{eq:214}) into the resonant condition at
anomalous Doppler effect and assuming $\delta \gg 1/2\gamma_{res}^2$
we obtain for the resonant frequencies
\begin{equation}
\omega_{1,2, res}\approx \frac{\omega_B}{\delta \gamma_{res}} \mp
\Delta\gamma \omega_B. \label{eq:33}
\end{equation}

The case $\Upsilon \ll 1$ was considered in \citet{KMM91}. In this
limiting case according to the last equation left and right handed
circularly polarized waves are generated by cyclotron instability
with approximately same frequencies and growth rates. In the
framework of this model the sign reversal of the circular
polarization, frequently observed near the centre of mean pulse
profile, is caused by drift motion of electrons and positrons in
the curved magnetic field of pulsar.

\subsection{The case $\Upsilon > 1$}

In the case $\Upsilon > 1$, right-handed circularly polarized wave
become super-luminal and consequently the resonant condition at
anomalous Doppler effect can not be fulfilled for this wave mode.
Therefore only left-handed circularly polarized waves are generated
by cyclotron instability. In the framework of the presented model
this situation is realized in the pulsars that have strong
domination of one of the circularly polarized component along the
whole averaged pulse profile for all observed frequencies, such as
PSR B1913+10, PSR B1914+13 and PSR B1356-60 \citep{HMXQ98}. This
suggestion is supported by the fact, that the pulsars of this group
as a rule have high or extremely high degree of circular
polarization ($\langle |V |\rangle/S >15 \% $). Indeed, as it was
mentioned in the previous section, circularly polarized waves exist
only in the small angle $\theta_0$ with respect to the pulsar
magnetic field. According to equation (\ref{eq:212}) the value of
$\theta_0$ as well as $\Upsilon$ is proportional to $\Delta \gamma$,
and therefore proposed model predicts that pulsars with strong
domination of one circularly polarized component should have
relatively high degree of circular polarization.

\subsection{The case $0.1 < \Upsilon < 1$}

We suggest that the pulsars that shows strong frequency dependence
of the circular polarization has $0.1 < \Upsilon < 1$ in the wave
generation region. In this range of parameters waves with both left
and right-handed circularly polarized waves are generated by
cyclotron instability, but there exists significant difference
between resonant frequencies given by equation (\ref{eq:33}) and
corresponding growth rates (\ref{eq:216}). One also has to note,
that the resonant frequencies as well as the decrements vary with
distance from the pulsar surface. Therefore, in general quite
complicated picture of circular polarization dependence on frequency
can be supposed. However, the qualitative picture of the frequency
dependence of radio emission circular polarization of such pulsars
should be the following: in case of positive $\Delta\gamma$, at low
and high frequencies right-handed and left-handed circular
polarization should strongly dominate, whereas in the range of
middle frequencies some soft transition between these two regimes
should be observable. This qualitative picture is in good accordance
with observations of pulsars that shows strong frequency dependence
of circular polarization, e.g., PSR B1749-28 and PSR B1240-64.
According to the presented model, due to the reasons discussed in
the previous subsection, these pulsars also should have relatively
high degree of circular polarization. It seems that, existing
observations confirm this feature \citep{HMXQ98}.

It should be noted that analysis presented in the beginning of this
section shows that $\Delta\gamma/\gamma_p$ is mainly determined by
two unmeasurable variables: the ratio $R_c/R_0$ and the ratio of
dipole and quadrupole components of the magnetic field at the
surface of the pulsar, whereas its dependence on measurable
quantities, such as $P$ and $\dot P$ is very weak. Consequently,
direct comparison with observational data seems to be very difficult
and only indirect analysis presented above could be performed for
qualitative comparison of the predictions of the presented model
with observations.

\section{Conclusions}

In this paper we attempt to explain observed properties of the
circular polarization in radio emission of pulsars in the framework
of the model that involves generation of circularly polarized waves
in the magnetosphere by cyclotron instability. Relative motion of
electrons and positrons in the magnetosphere plasma is responsible
for existence of tho circularly polarized waves that are generated
by cyclotron instability. According to resonant condition at
anomalous Doppler effect, the critical parameter that determines the
character of cyclotron instability development is $\Upsilon$ defined
by equation (\ref{eq:32}). If $\Upsilon > 1$, one of the waves
becomes super-luminal and therefore can not be generated by
cyclotron instability. In our opinion this case is realized in the
pulsars that have strong domination of one of the circularly
polarized component along the whole averaged pulse profile for all
observed frequencies. When $0.1 < \Upsilon < 1$, waves with both
left and right-handed circular polarization are effectively
generated, but there exists significant difference between resonant
frequencies given by equation (\ref{eq:33}). This corresponds to the
pulsars with strong frequency dependence of circular polarization
profiles. And finally, the case $\Upsilon \ll 1$ is realized in
pulsars with asymmetric circular polarization profiles.

Explanation of observed circular polarization of pulsars should
include both generation and propagation effects. While in the
present paper we focus on generation effects, the propagation
effects of the waves in the magnetosphere plasma will be described
in the consequent paper. relevant studies are under way presently.

\bsp

\label{lastpage}

\end{document}